\documentclass[%
 aip,
 amsmath,amssymb,
 reprint,%
]{revtex4-1}

\usepackage{graphicx}% Include figure files
\usepackage{dcolumn}% Align table columns on decimal point
\usepackage{bm}% bold math
\usepackage[utf8]{inputenc}
\usepackage[T1]{fontenc}
\usepackage{mathptmx}

\begin{document}

\preprint{AIP/123-QED}

\title[Influence of Furnace Baking on Q-E Behavior of Superconducting Accelerating Cavities]{Influence of Furnace Baking on Q-E Behavior of Superconducting Accelerating Cavities}%
%\title{New Baking Process on Superconducting Accelerating Cavities}%

\author{H. Ito}
\email{hayato.ito@kek.jp}
\affiliation{ 
High Energy Accelerator Research Organization (KEK), 305-0801 Tsukuba, Ibaraki, Japan%\\This line break forced with \textbackslash\textbackslash
}%

\author{H. Araki}%
\affiliation{ 
High Energy Accelerator Research Organization (KEK), 305-0801 Tsukuba, Ibaraki, Japan%\\This line break forced with \textbackslash\textbackslash
}%

\author{K. Takahashi}%
\affiliation{The Graduate University for Advanced Studies, SOKENDAI, 305-0801 Tsukuba, Ibaraki, Japan%\\This line break forced with \textbackslash\textbackslash
}%

\author{K. Umemori}%
\affiliation{ 
High Energy Accelerator Research Organization (KEK), 305-0801 Tsukuba, Ibaraki, Japan%\\This line break forced with \textbackslash\textbackslash
}%
\affiliation{The Graduate University for Advanced Studies, SOKENDAI, 305-0801 Tsukuba, Ibaraki, Japan%\\This line break forced with \textbackslash\textbackslash
}%
\date{\today}% It is always \today, today,

\begin{abstract}
The performance of superconducting radio-frequency (SRF) cavities depends on the niobium surface condition.
Recently, various heat-treatment methods have been investigated to achieve unprecedented high quality factor (Q) and high accelerating field (E). 
We report the influence of a new baking process called furnace baking on the Q-E behavior of 1.3 GHz SRF cavities.
Furnace baking is performed as the final step of the cavity surface treatment; the cavities are heated in a vacuum furnace for 3 h, followed by high-pressure rinsing and radio-frequency measurement. 
This method is simpler and potentially more reliable than previously reported heat-treatment methods, and it is therefore, easier to apply to the SRF cavities.
We find that the quality factor is increased after furnace baking at temperatures ranging from 300$^\circ$C to 400$^\circ$C, while strong decreasing the quality factor at high accelerating field is observed after furnace baking at temperatures ranging from 600$^\circ$C to 800$^\circ$C. 
We find significant differences in the surface resistance for various processing temperatures. 
\end{abstract}

\maketitle

A superconducting radio-frequency (SRF) cavity is a key component of particle accelerators used to generate charged particle beams.
An SRF cavity exhibits a lower energy dissipation and a lower surface resistance ($R_{\rm s}$) under a radio frequency (RF) field, compared to a normal-conducting accelerating cavity, which enables continuous-wave operation at a high accelerating field ($E_{\rm acc}$).
Owing to decades of research focused on the improvement of SRF cavities \cite{Padamsee2017, Gurevich2017}, various surface treatment techniques have been established; thus SRF cavities with superior performance in terms of the quality factor ($Q_{0}$) and $E_{\rm acc}$ have been developed \cite{rongli, watanabe, Kubo:IPAC2014}.

In recent years, further surface treatment techniques, such as nitrogen doping \cite{GrassellinoNdope, PashupatiN-dope-infushon}, nitrogen infusion \cite{GrassellinoNinfusion, PashupatiN-dope-infushon}, and two-step baking \cite{GrassellinoTwostep}, have been investigated to increase the $Q_{0}$ and $E_{\rm acc}$.
The nitrogen-doped cavities have an extremely high $Q_{0}$ and show increasing the $Q_{0}$ as a function of the $E_{\rm acc}$ which is referred to as the anti-Q slope. 
However, the maximum $E_{\rm acc}$ obtained with nitrogen doping is lower than that of the conventional surface-treated cavity.
Moreover, the nitrogen-doped cavities are highly sensitive to trapped magnetic flux compared with standard treated cavities \cite{MartinelloSensitivity}.
The nitrogen doping process has been applied in the Linac Coherent Light Source (LCLS$\rm - I\hspace{-.1em}I$) cavity fabrication process \cite{BishopLCLS, GonnellaLCLS} because it is highly reproducible and has resulted in high $Q_{0}$ and anti-Q slope in several studies \cite{PashupatiN-dope-infushon}.
In the nitrogen infusion technique, the Q-E behavior does not change significantly, and both $Q_{0}$ and $E_{\rm acc}$ are improved compared with those values obtained using the standard surface treatment methods \cite{PashupatiNinfusion,UmemoriNinfusion,WenskatNinfusion}; however, the reproducibility has been limited to a few laboratories.
The two-step baking process developed at Fermi National Accelerator Laboratory (FNAL) can produce cavities with a maximum $E_{\rm acc}$ of approximately 50 MV/m \cite{GrassellinoTwostep,bafiaTwostep}, and other laboratories are currently verifying the effectiveness of two-step baking.

In the typical surface treatment planned for the International Linear Collider (ILC), the following procedure is implemented: after fabricating the SRF cavity, a 100 $\mu$m layer of the cavity inner surface is removed by bulk electropolishing; this results in the elimination of the surface layer damaged during cavity fabrication.
After electropolishing, the surface is thoroughly rinsed with ultrapure water, and ultrasonic cleaning is performed by filling ultrapure water with a surfactant, followed by high-pressure ultrapure water rinsing (HPR). 
Then, annealing is performed in a vacuum furnace at approximately 800$^\circ$C to desorb the hydrogen that was absorbed on the niobium surface during electropolishing.
Subsequently, light electropolishing is used to remove a layer of approximately 20 $\mu$m of the cavity inner surface to eliminate dirt from the inner surface, followed by sufficient water rinsing, ultrasonic cleaning, HPR, and assembly in a cleanroom.
Next, as the final step in the surface treatment process, the cavity is vacuumed and heat-treated at 120$^\circ$C for 48 h.

In this study, a new heat-treatment method, which is simpler and more reliable than the surface treatment method described above, is investigated from the viewpoint of oxygen diffusion of a niobium oxide layer, and the effects on the properties of $Q_{0}$, the Bardeen-Cooper-Schrieffer (BCS) resistance ($R_{\rm BCS}$), and the residual resistance ($R_{\rm res}$) for each $E_{\rm acc}$ are studied.
In the 1980s and 1990s, SRF cavities that were heat-treated in vacuum in the range of 250 to 300$^\circ$C and 1100 to 1400$^\circ$C were investigated to understand the effect of an oxide layer on the SRF cavity performance \cite{PalmerMid-T,PalmerMid-T2,PalmerHigh-T}.
It was revealed that the heat-treatment at 250$^\circ$C dissolves the oxide layer and decreases $R_{\rm BCS}$.
Therefore, it is expected that the heat treatment in this study can be performed in the same temperature range to create a cavity with high $Q_{0}$.
We used several 1.3 GHz TESLA and STF (TESLA-like) single-cell cavities that had undergone various surface treatments.
As a first step, a 10 or 20 $\mu$m layer of the cavity inner surface was removed by the light electropolishing to reset the surface conditions in the cavity, followed by HPR to eliminate any remaining impurities on the surface.
Subsequently, the cavities were placed in a large vacuum furnace.
The inner diameter of the furnace chamber is $\phi$ 950 mm, and the length is 2080 mm \cite{UmemoriLinac18}.
%All exhaust systems consist of oil-free pumps.
This vacuum furnace can be depressurized to $1 \times 10^{-6}$ Pa at room temperature using a cryopump.
Due to the insufficient cooling capacity of the cryopump, heat treatment at 800$^\circ$C increases the temperature of the cryopump. In such a case, we switch from the cryopump to a turbo molecular pump for achieving the desired level of vacuum \cite{UmemoriLinac18}.
%Switching to TMP increases the vacuum by one order.
The Quadrupole Mass Spectrometer (Q-mass) was equipped with a vacuum furnace to monitor the partial pressure of each element during heat treatment.
The cavities were baked in a temperature range of 200 to 800$^\circ$C for 3 h in this vacuum furnace.
This baking process is referred to as ``furnace baking'', which is different from the ``medium-temperature bake'' (mid-T bake) that is performed at FNAL and Institute of High Energy Physics (IHEP) \cite{PosenMid-T, ZhouMid-T}.
The mid-T bake process requires special heat treatment equipment to perform the RF measurement without exposing the inner surface of the cavity to the air after heat treatment at 250 to 400$^\circ$C, whereas the furnace baking process is a simple method that can be performed with existing cavity treatment systems because the heat treatment is performed in a vacuum furnace.
In the first step, the baking time was fixed at 3 h to investigate the optimum temperature and achieve high $Q_{0}$.
To perform the furnace baking, the cavity temperature was ramped up from room temperature at a ramp rate of 200$^\circ$C/h to a target temperature of 200 to 800$^\circ$C. 
The vacuum was 1-2$\times 10^{-6}$ Pa at room temperature, and it was maintained at the order of $10^{-4}$ Pa even during baking.
After the furnace baking process and cooling down to below 50$^\circ$C, the furnace was purged with $\rm N_{2}$ gas, and the cavity was packed and placed on the HPR stand for final rinsing before assembly.
A new niobium oxide layer grows on the inner surface during this step because of exposure to air and water; however, this is not expected to affect the oxygen diffusion region formed by furnace baking.
After assembly, no further baking was performed, and the RF measurements were conducted.

%High-pressure ultrapure water from pipes placed inside the cavity impacts the inner surface of the cavity, and the cavity rotates up and down to clean the entire inner surface of the cavity.
%With the flange of the top beam tube open, the inner surface HPR is performed for 1.5 hours, followed by 15 minutes of outer surface HPR with the flange of the top beam tube closed. Finally, the HPR is completed by performing the inner surface HPR again for 1.5 hours with the flange of the top beam tube closed.
%The cavity goes up and down 6 times in 1.5 hours of HPR.
%The pressure of the water in the pipe is 8 MPa.

The cavity was then cooled down by depressurizing liquid helium to 1.5 K, which is the lowest temperature that can be achieved in the cryostat available at High Energy Accelerator Research Organization (KEK).
Then, the Q-E curve at each temperature was obtained by calculating the input, reflected, and transmitted RF powers for 0.1 K temperature increments. Finally, the RF measurement was performed up to the quench field at 2 K.
To minimize the magnetic flux trapping during the cooling process, the magnetic field around the cavity was reduced to less than $\sim$1 mG using the magnetic shield and solenoid coil, and a heater placed at the top beam tube was used to provide a temperature gradient in the cavity for flux expulsion \cite{HuangFluxExpulsion, PosenFluxExpulsion, PashupatiFluxExpulsion}.
Figure 1 shows the Q-E curve for the cavity that was furnace-baked at 350$^\circ$C.
The cavity was quenched once during the measurement at 1.5 K, which caused it to trap the magnetic flux and subsequently decrease the $Q_{0}$. However, higher $Q_{0}$ and anti-Q slope were still clearly observed in the 2 K measurement results compared to the standard treated cavity (see Fig. 3).

\begin{figure}[h]
\includegraphics[clip, width=8.5cm]{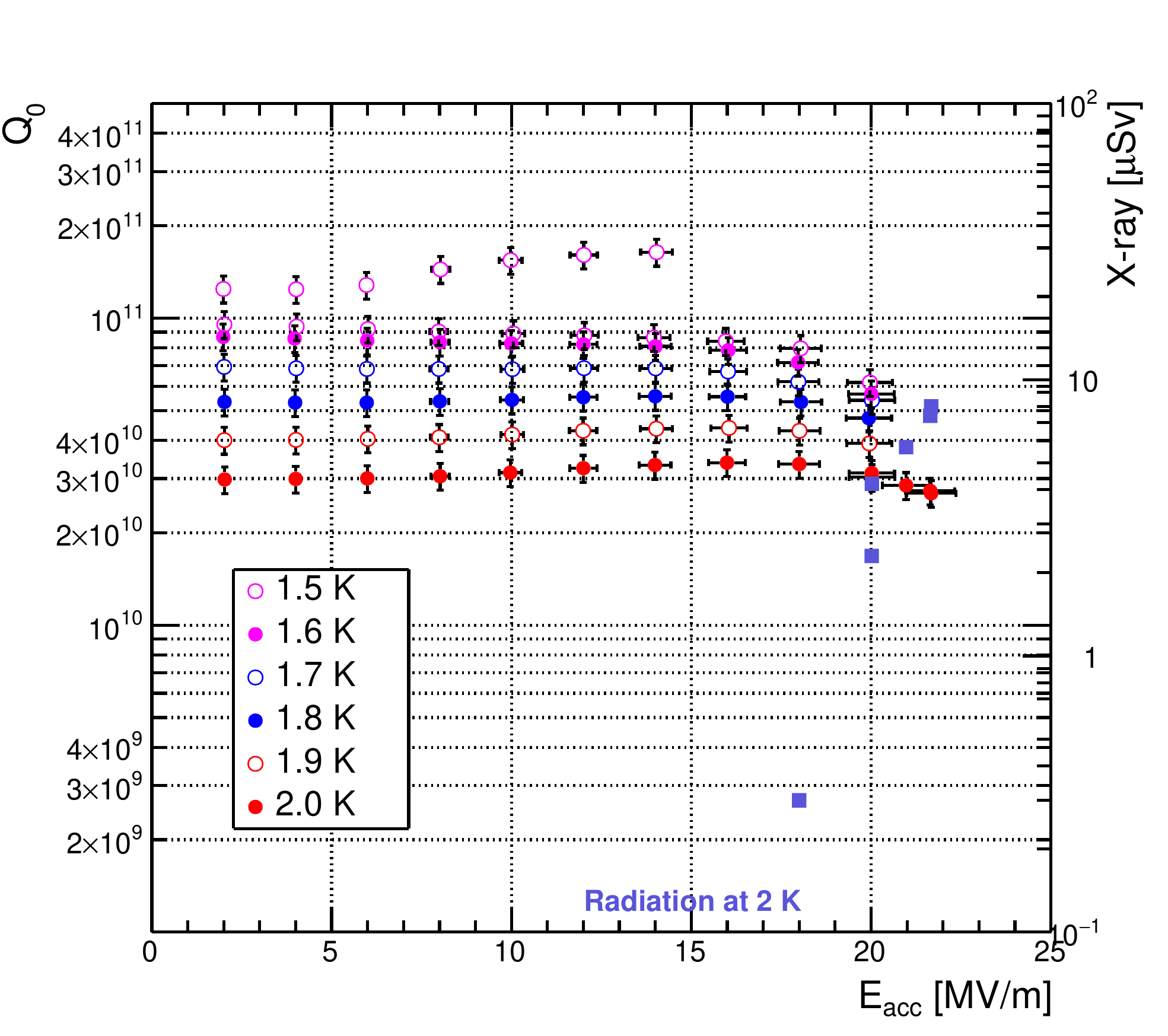}% Here is how to import EPS art
\caption{\label{fig:QE350} Q-E curve at each temperature for the cavity that was furnace-baked at 350$^\circ$C. Colored circles show Q-E curves at each temperature, and purple squares show radiation levels at 2 K measurement.}
\end{figure}

%No such event was observed in the other measured cavities.
$R_{\rm s}$ at each temperature and $E_{\rm acc}$ is calculated using $R_{\rm s} = G / Q_{0}$, where G is the geometric factor that is independent of material properties \cite{padamsee}.
$R_{\rm s}$ can be expressed as the sum of $R_{\rm BCS}$, which decreases exponentially with temperature, and $R_{\rm res}$, which is a weak temperature-dependent or temperature-independent term that cannot be accounted for in $R_{\rm BCS}$.
$R_{\rm s}$ is decomposed into $R_{\rm BCS}$ and $R_{\rm res}$ at each $E_{\rm acc}$ using the data set at the same $E_{\rm acc}$ from the Q-E curve.
The decomposition is performed using the following fitting equation:
\begin{equation}
 R_{\rm s}(T) = \frac{A \omega ^2}{T} e^{-(\Delta/kT)} + R_{\rm res},
\end{equation}
where A is a fitting constant that depends on superconducting properties, T is the temperature, $k$ is the Boltzmann constant, 2$\Delta$ is the energy gap of the superconductor, which is treated as a fitting parameter, and $\omega$ is the frequency of the cavity.
The first term in this equation corresponds to $R_{\rm BCS}$.
The colored curves in the upper figure of Fig. 2 show fitting parameters for $R_{\rm s}(T)$ at each $E_{\rm acc}$.
The red closed circles in the lower figure of Fig. 2 illustrate the behavior of $R_{\rm res}$ for $E_{\rm acc}$.
The blue closed circles depict the behavior of $R_{\rm BCS}$, which decreases sharply as $E_{\rm acc}$ increases in the case of 350$^\circ$C furnace baking.
This behavior is considerably different from the behavior of standard treated cavities (120$^\circ$C and 48 h baked cavities), and it is similar to the behavior observed in a nitrogen-doped cavity \cite{GrassellinoNdope}.
Red open circles depict the estimated behavior of $R_{\rm res}$ before the flux trapping; $R_{\rm res}$ is smaller than that of the standard treated cavities for 350$^\circ$C furnace baking.

\begin{figure}[h]
 \begin{minipage}{1\hsize}
  \begin{center}
   \includegraphics[width=8cm]{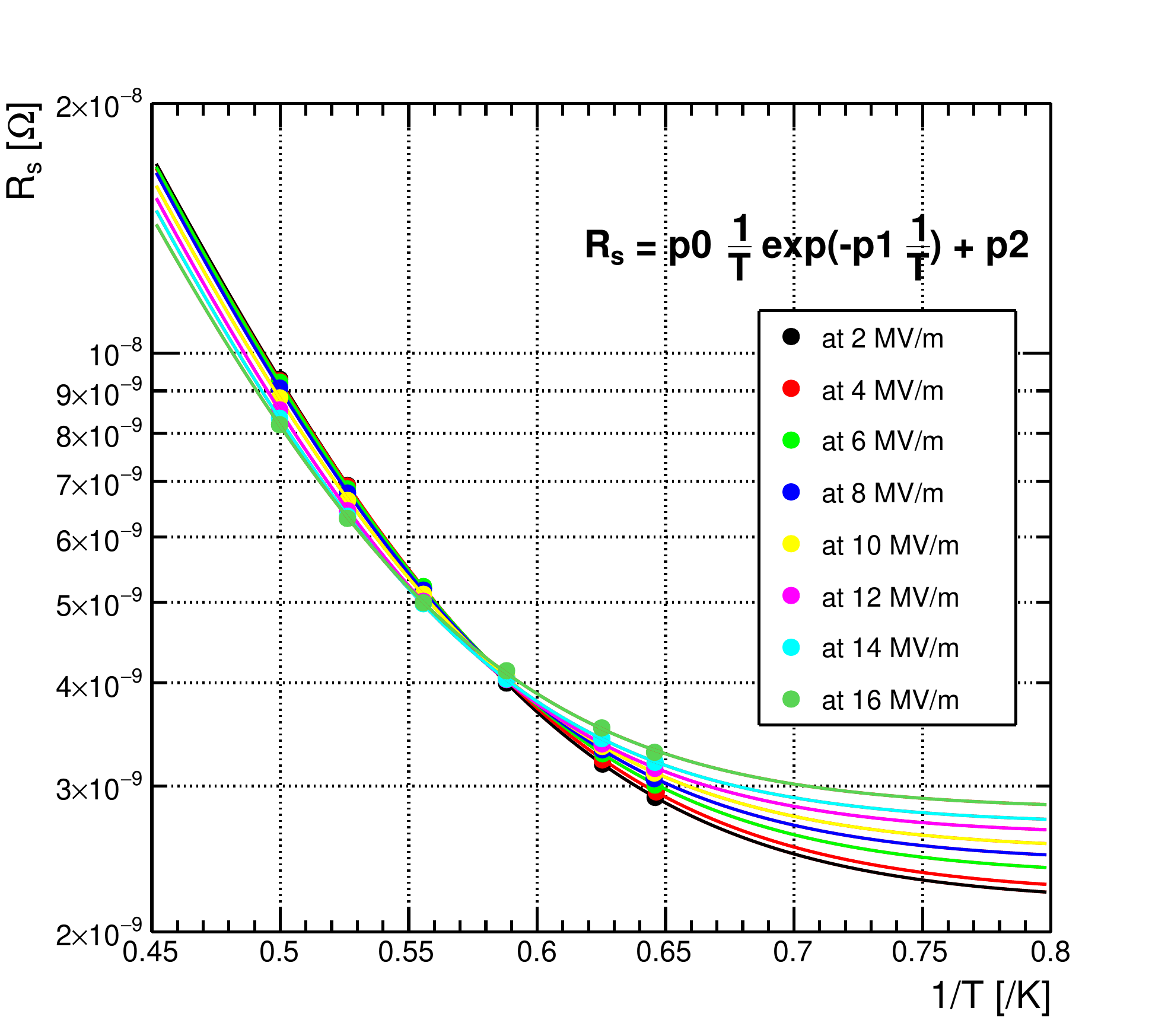}
  \end{center}
  \label{fig:one}
 \end{minipage}\\
 \begin{minipage}{1\hsize}
  \begin{center}
   \includegraphics[width=8cm]{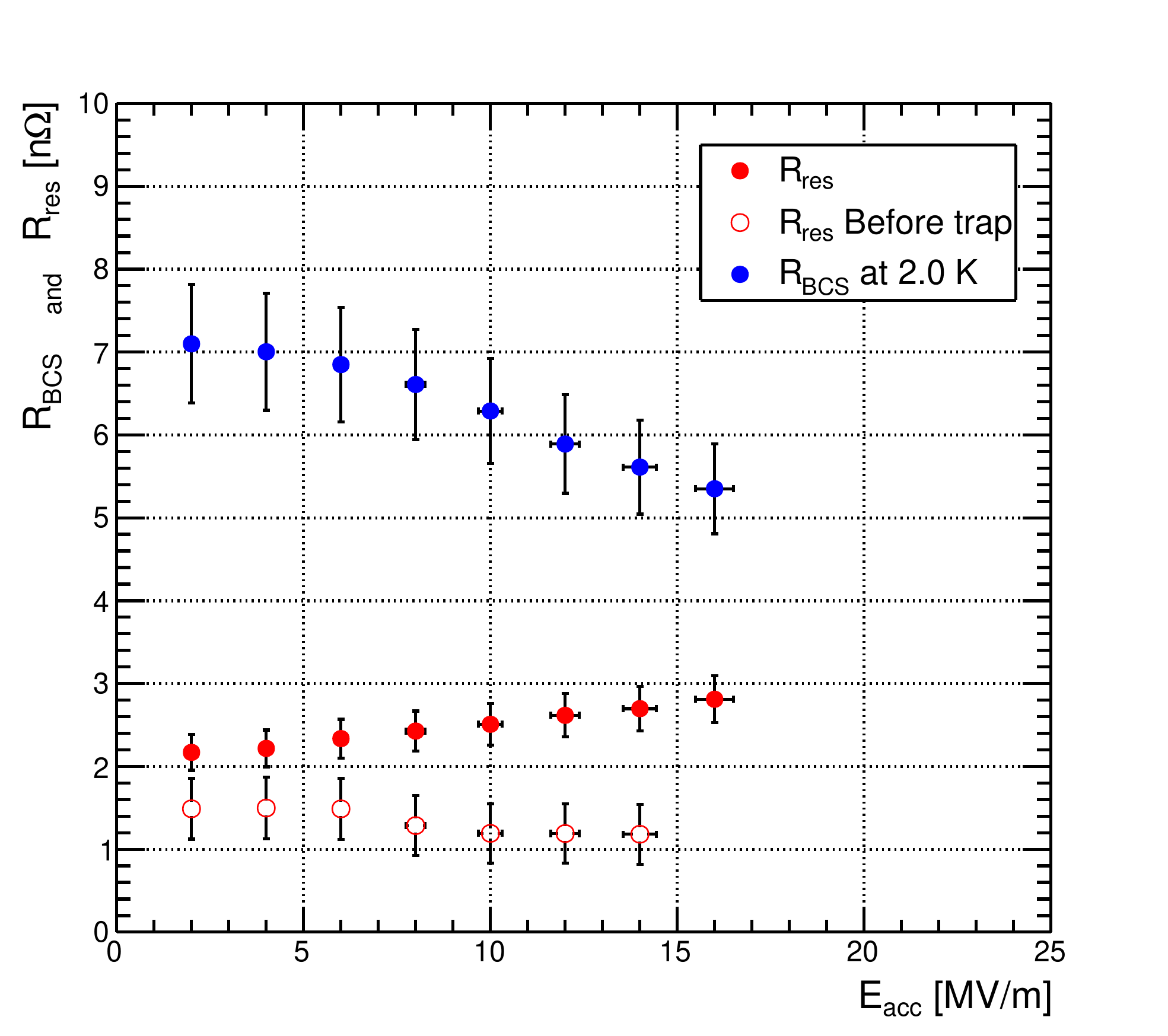}
  \end{center}
  \label{fig:two}
 \end{minipage}
  \caption{Temperature dependence of $R_{\rm s}$ for each $E_{\rm acc}$ (upper figure) and behavior of $R_{\rm BCS}$ and $R_{\rm res}$ for $E_{\rm acc}$ (lower figure). $R_{\rm res}$ before the trap was obtained by subtracting $R_{\rm BCS}$ obtained after the trap from the $R_{\rm s}$ at 1.5 K before the trap.}
\end{figure}

Figure 3 shows a comparison of the Q-E curves measured at 2 K for cavities that were furnace-baked at various temperatures (200 to 800$^\circ$C) and a standard treated cavity (120$^\circ$C and 48 h baking under vacuum directly followed by RF measurement, no exposure to air or water).
The Q-E behavior of the 200$^\circ$C and 3 h furnace-baked cavity (purple points) is similar to that of the standard treated cavity (black points).
This indicates that the conventional performance can be achieved by replacing the 120$^\circ$C and 48 h baking with 200$^\circ$C and 3 h furnace baking, which may be significantly effective for mass production of cavities.
Four furnace-baked cavities, baked at 300 to 400$^\circ$C, have high $Q_{0}$ and anti-Q slope, but a low $E_{\rm acc}$ compared with the standard treated cavity.
This behavior is typically associated with nitrogen-doped cavities \cite{GrassellinoNdope}.
In particular, 300$^\circ$C furnace baking produces an extremely high Q cavity, with a $Q_{0}$ of over 5$\times 10^{10}$ at 16 MV/m.
Furthermore, 300$^\circ$C furnace baking has the same effect for two different cavities, indicating good reproducibility.
These results are in good agreement with those obtained for single-cell cavities that are furnace-baked in the temperature range of 250 to 400$^\circ$C at IHEP \cite{FeisiFurnacebaking}.
The high-temperature furnace-baked cavities baked at 600$^\circ$C and 800$^\circ$C did not reach high $Q_{0}$, and the Q values were comparable to those of the standard treated cavity.
A phenomenon called high field Q slope (HFQS), in which the $Q_{0}$ decreases significantly at high $E_{\rm acc}$ \cite{safaSRF01}, was observed in these cavities.
This HFQS is considered to be related to the diffusion of oxygen and hydrogen on the inner surface of the cavity \cite{BenvenutiSRF01, CiovatiDiffusion, CiovatiHFQS, RomanenkoHFQS, ChecchinHFQS}, and the effect of suppressing the HFQS diminished due to oxygen diffusion at high temperatures.
These results indicate that diverse Q-E behaviors were obtained when the furnace baking temperature was changed from 200 to 800$^\circ$C.
In particular, furnace baking at 300 to 400$^\circ$C resulted in high $Q_{0}$ and anti-Q slope.
The low $E_{\rm acc}$ is similar to that of the nitrogen-doped cavity and may be related to the low superheating field at the dirty limit\cite{KuboHsh}.
Theoretical considerations suggest that the $Q_{0}$ varies depending on the cavity surface condition\cite{GurevichSurface, KuboField-dependent, KuboWeak-field}. 

\begin{figure}[h]
\includegraphics[clip, width=8.5cm]{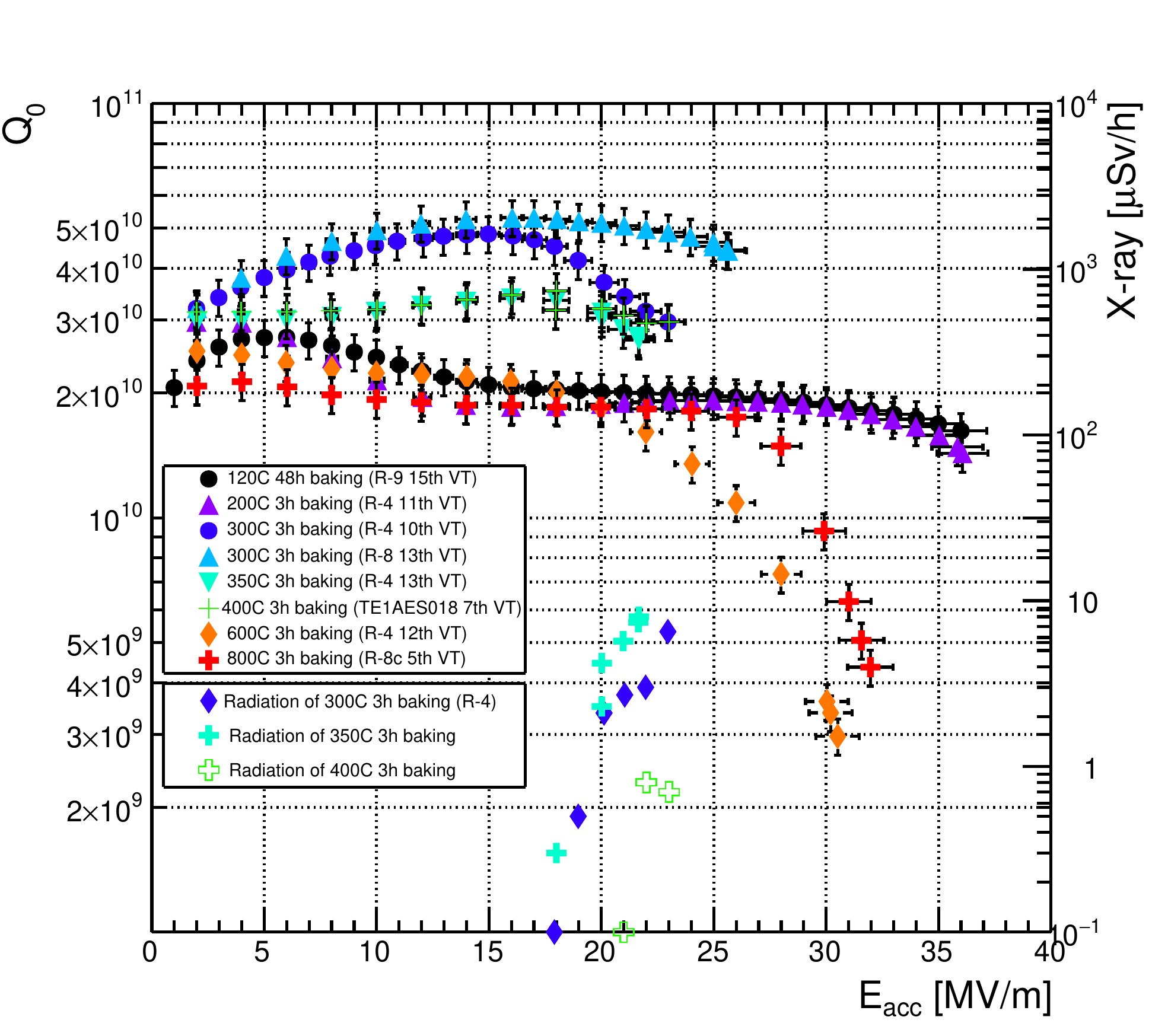}% Here is how to import EPS art
\caption{\label{fig:QE} Comparison of Q-E behavior measured at 2 K for cavities that were furnace-baked at various temperatures (200 to 800$^\circ$C) and a standard treated cavity (120$^\circ$C and 48 h baking). Colored closed points show the Q-E curve for each furnace-baked cavity, and the colored opened points show the radiation levels corresponding to each colored closed point.}
\end{figure}

The upper panel of Fig. 4 shows the relationship between $R_{\rm BCS}$ at 2 K and $E_{\rm acc}$ for each furnace-baked and standard treated cavity.
$R_{\rm BCS}$ behavior is classified into three types: one that increases with increasing $E_{\rm acc}$, one that decreases with increasing $E_{\rm acc}$, and one that does not increase as much as the first type.
The 200$^\circ$C furnace-baked cavity and the standard treated cavity correspond to the first type mentioned above, and the slope of $R_{\rm BCS}$ is steep compared with those obtained for other cavities.
The 300 to 400$^\circ$C furnace-baked cavities correspond to the second type.
In these cavities, $R_{\rm BCS}$ decreases as $E_{\rm acc}$ increases, which is the origin of the anti-Q slope.
This behavior is the most pronounced in 300$^\circ$C furnace-baked cavities.
The 600$^\circ$C and 800$^\circ$C furnace-baked cavities correspond to the third type.
These cavities already have a high $R_{\rm BCS}$ at low $E_{\rm acc}$; however, the slope is less steep compared with the first type.
From these results, it was found that $R_{\rm BCS}$ behavior varies significantly with differences in the baking temperature, resulting in the variation of Q-E behavior.
Further, it was suggested that there is an inflection point between 200$^\circ$C and 300$^\circ$C, where the behavior of $R_{\rm BCS}$ changes significantly, and that a similar inflection point exists in the region between 400$^\circ$C and 600$^\circ$C.
The lower panel of Fig. 4 shows the relationship between $R_{\rm res}$ and $E_{\rm acc}$ for each furnace-baked cavity and standard treated cavity.
It was found that $R_{\rm res}$ is lower for all the furnace-baked cavities compared with that of the standard treated cavity, and $R_{\rm res}$ behavior changes with differences in baking temperature but not as drastically as $R_{\rm BCS}$ behavior.
Notably, the 600$^\circ$C furnace-baked cavity has an extremely low $R_{\rm res}$ of 0.2 $\rm n \Omega$, which corresponds to a $Q_{0}$ of over $1 \times 10^{12}$.
Because $R_{\rm res}$ dominates $R_{\rm s}$ at temperatures of approximately 1 K, this 600$^\circ$C furnace baking has the potential to be a useful processing method for superconducting devices, such as those used at cryogenic temperatures in the mK region rather than the SRF accelerator application operating at 2 K.

\begin{figure}[h]
 \begin{minipage}{1\hsize}
  \begin{center}
   \includegraphics[width=8.5cm]{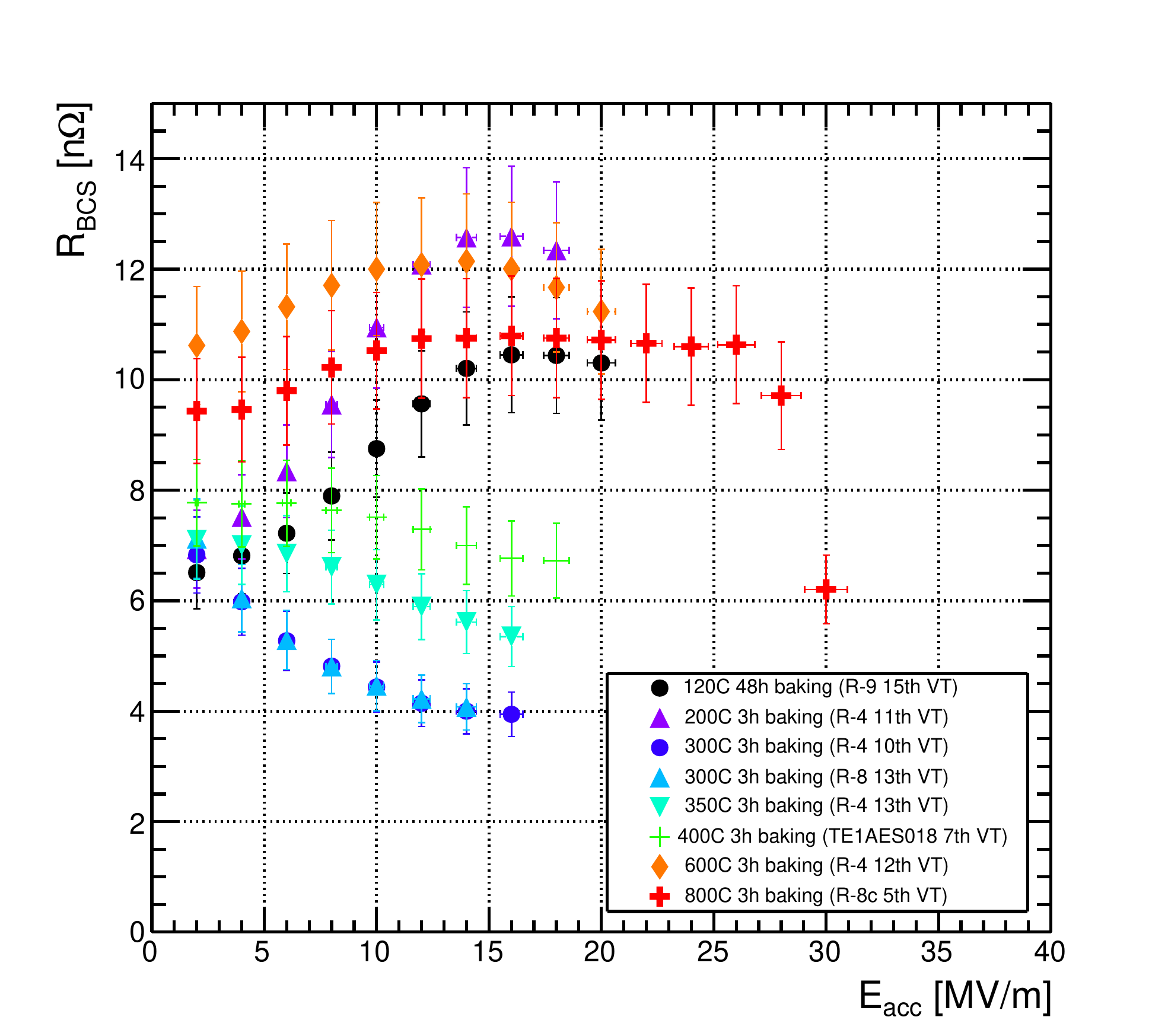}
  \end{center}
  \label{fig:RBCS}
 \end{minipage}\\
 \begin{minipage}{1\hsize}
  \begin{center}
   \includegraphics[width=8.5cm]{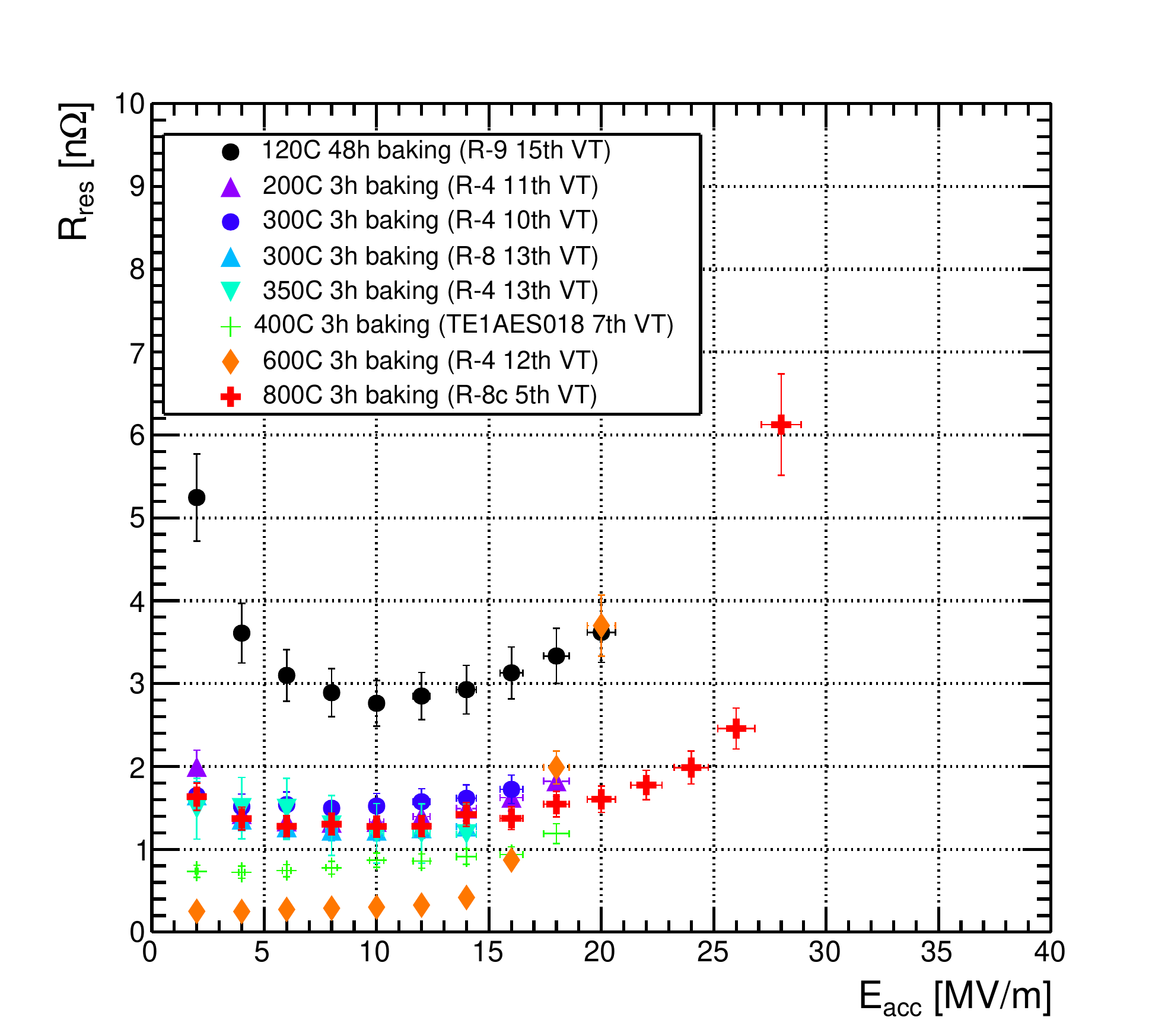}
  \end{center}
  \label{fig:RRes}
 \end{minipage}
  \caption{$R_{\rm BCS}$ behavior at 2 K for $E_{\rm acc}$ for each furnace-baked cavity and standard treated cavity (upper panel). Relationship between $R_{\rm res}$ and $E_{\rm acc}$ for each furnace-baked cavity and standard treated cavity (lower panel).}
\end{figure}

The sensitivity of the mid-T (300 to 400$^\circ$C) furnace-baked cavity was estimated by measuring the Q-E curve after cooling slowly in a 20 mG field. (Slow cooling allowed the magnetic field of 95\% to be trapped in the cavity.)
The sensitivity $S$ describes the amount of increase in $R_{\rm s}$ per unit of trapped field $B_{\rm trap}$ and can be expressed as
\begin{equation}
 S= \frac{\Delta R_{\rm s}}{B_{\rm trap}}
.
\end{equation}
Figure 5 shows the measurement results of the sensitivity of the mid-T furnace-baked cavity and the comparison to a standard treated cavity.
The mid-T furnace-baked cavities have a high sensitivity compared with the standard treated cavity and resemble the nitrogen-doped cavity \cite{MartinelloSensitivity}.
The sensitivity of the 300$^\circ$C furnace-baked cavity is higher than that of the nitrogen-doped cavity.
When such a high-sensitivity cavity is installed as an accelerator component, it is necessary to consider the effect of the magnetic field trapping due to the ambient magnetic field more severely than in the standard treated cavity because it is difficult to realize the ideal magnetic field shielding in real accelerator components as in the RF measurement of this study.

\begin{figure}[h]
\includegraphics[clip, width=8.5cm]{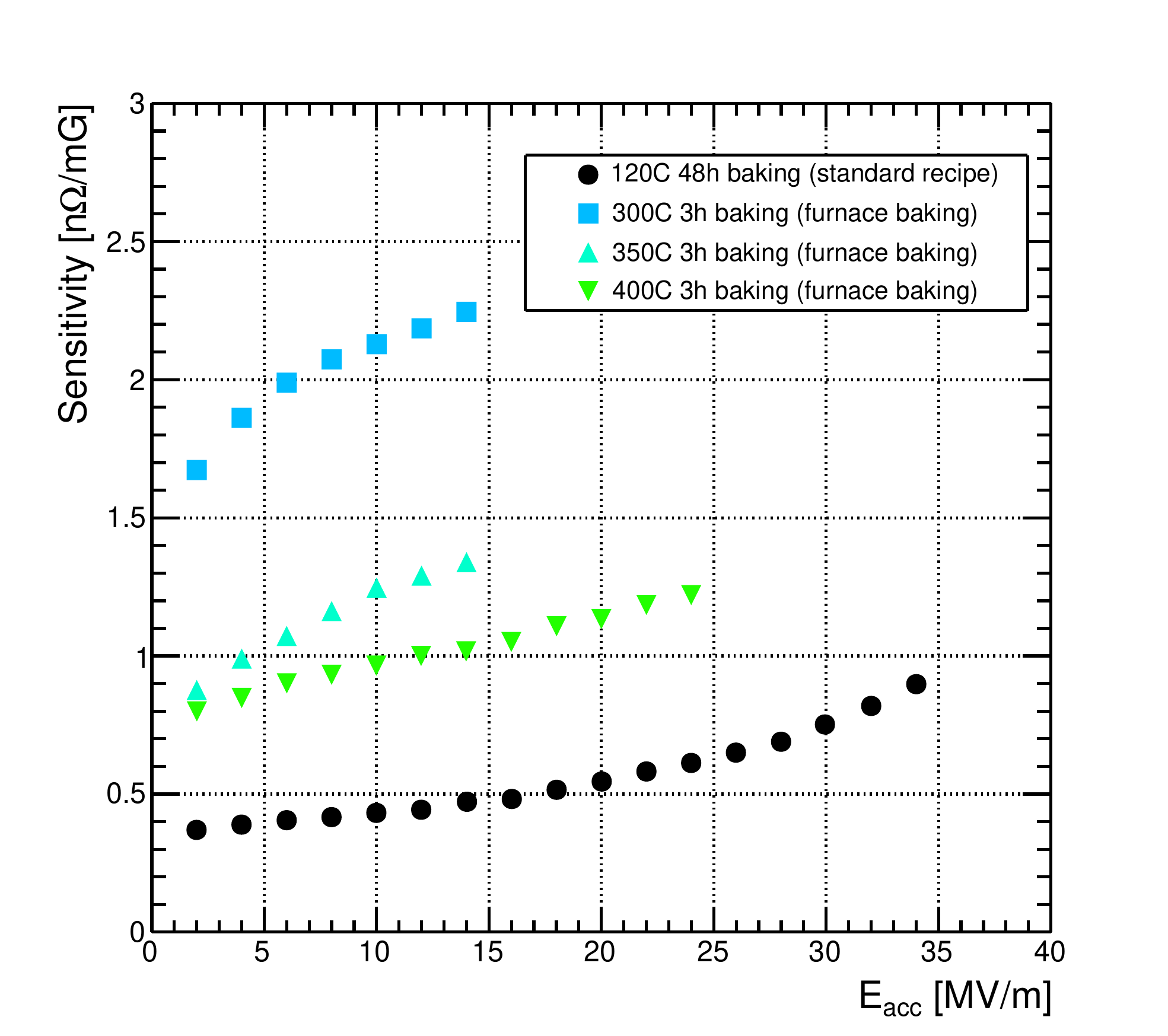}% Here is how to import EPS art
\caption{\label{fig:Sensitivity} Sensitivity of mid-T furnace-baked cavities and standard treated cavity.}
\end{figure}

In this study, a new heat-treatment process called furnace baking has been investigated at several baking temperatures ranging from 200 to 800$^\circ$C and a baking time of 3 h.
The behavior of Q-E is found to be sensitive to the baking temperature.
Furthermore, the quench field changes with the baking temperature.
The mid-T (300 to 400$^\circ$C) furnace baking produces a cavity with high $Q_{0}$ and anti-Q slope.
In particular, the 300$^\circ$C furnace-baked cavity has an extremely high $Q_{0}$ of over 5$\times 10^{10}$ at 16 MV/m and 2 K.
The quench field is 20 to 25 MV/m for the mid-T furnace-baked cavities, which is lower than the quench field of the standard treated cavity.
Although the achievable $E_{\rm acc}$ is low, its high $Q_{0}$ is very impressive, and combined with the simplicity of the furnace baking procedure, it is clear that the mid-T furnace baking can be successfully adapted to various SRF applications in the future.
$R_{\rm res}$ is lower for all the furnace-baked cavities compared with that of the standard treated cavity.
For the 600$^\circ$C furnace-baked cavity, an extremely low $R_{\rm res}$ of 0.2 $\rm n \Omega$ is obtained.
The sensitivity of the mid-T furnace-baked cavity is higher than that of the standard treated cavity and resembles that of the nitrogen-doped cavity.
The 300$^\circ$C furnace-baked cavity, which has the highest $Q_{0}$, has a higher sensitivity than that of the nitrogen-doped cavity.
Further studies will be undertaken by focusing on the surface analysis based on the sample study to reveal the relationship between these behaviors and the cavity surface condition.
Further, process optimization will be performed by changing the baking time.

This work was supported by JSPS Grant-in-Aid for Scientific Research(B) No. 19H04402.

\nocite{*}
\bibliography{references}
\end{document}